\documentclass[twocolumn,showpacs,preprintnumbers,amsmath,amssymb]{revtex4}

\usepackage{graphicx}
\usepackage{dcolumn}
\usepackage{bm}

\begin{document}


\title{Collimated, single-pass atom source from a pulsed alkali
metal dispenser for laser-cooling experiments}

\author{Kevin L. Moore}
\email{klmoore@socrates.berkeley.edu}
\author{Thomas P. Purdy}%
\author{Kater W. Murch}%
\author{Sabrina Leslie}%
\author{Subhadeep Gupta}%
\author{Dan M. Stamper-Kurn}%
\affiliation{%
University of California\\
366 LeConte Hall, Berkeley, CA  94720}%

\begin{abstract}
We have developed an improved scheme for loading atoms into a
magneto-optical trap (MOT) from a directed rubidium alkali metal
dispenser in $<$10$^{\text{-10}}$ torr ultra-high vacuum conditions.
A current-driven dispenser was surrounded with a cold absorbing
``shroud" held at $\leq$ 0 $^{\text{o}}$C, pumping rubidium atoms
not directed into the MOT.  This nearly eliminates background atoms
and reduces the detrimental rise in pressure normally associated
with these devices.  The system can be well-described as a
current-controlled, rapidly-switched, two-temperature thermal beam,
and was used to load a MOT with 3$\times 10^{8}$ atoms.
\end{abstract}

\maketitle

\section{\label{sec:level1}Introduction}

The first step in the construction of an atomic physics experiment
is obtaining an appropriate source of atoms.  \emph{Directed}
sources of atoms have a long and storied history \cite{ramsey:book}.
A thermal beam of atoms is easily obtained from an oven or other gas
source, though this inevitably involves a differential pumping
scheme, a 1/$r^{2}$ decrease in atom flux with distance between the
oven and the collection region, a mechanical shutter to quench the
beam, and direct handling of a purified sample of the atom of
interest. A Zeeman slower \cite{phillips:zeeman} can improve the
flux of laser-cooled atoms from an oven, but suffers from the same
drawbacks as an oven as well as the added complications of the
magnetic design and the slowing beams. A multiply loaded
magneto-optical trap (MOT) \cite{myatt:doubleMOT} initially loaded
from a vapor cell is a widely-used source for cold atom experiments,
but has the drawback of increased optical, electronic, and vacuum
infrastructure. Light-induced atomic desorption (LIAD)
\cite{anderson:LIAD}, while not a collimated beam source, is an
elegant technique which has recently been improved to yield very
fast MOT loading rates \cite{atutov:LIAD}. However, the
infrastructure necessary for fast LIAD is not always appropriate for
other experimental requirements.

Alkali metal dispensers \cite{SAES:getters}, or ``getters," have
emerged as a useful alternative to these sources
\cite{fortagh:getter, rapol:getter}, requiring only a modest
electric current ($<$10 A) for their operation.  The driving current
rapidly heats the dispenser causing a reduction reaction, inducing
the cm-scale devices to release an atomic vapor (rubidium, in our
case) with a rapid turn-off time. The emitted atoms are quite hot,
as the dispensers reach temperatures of 800 $^{\text{o}}$C or more
for typical current pulses \cite{fortagh:getter, rapol:getter}. The
fraction of atoms capable of being captured by a typical MOT is
quite small ($\approx$10$^{\text{-5}}$) due to the large
temperatures reached, but the efficacy of the dispensers for direct
loading is salvaged both by the large atom flux and their ability to
be placed close to the MOT. Alkali metal dispensers are already used
as sources for vapor cell MOTs (Refs \cite{lewandowski:thesis,
matthews:dynamical, wieman:getter} for example), but in these cases
the 800 $^{\text{o}}$C atoms are cooled by the walls of the vapor
cell so the loading rate into the MOT is increased.

Many ultracold atomic physics experiments demand base pressures of
less than 10$^{\text{-10}}$ torr, particularly in the case of
magnetic trapping of atoms for long periods of time.  Getters have
been used directly in an ultra-high vacuum (UHV) ultracold atom
experiment \cite{ott:BECchip} and the atomic flux from a dispenser
has been collimated to make an atom beam \cite{roach:getter}, but to
the best of the authors' knowledge no published work describes a
system which attempts to control the output flux of the getter
entirely in the UHV chamber.

The desire for a fast, simple, and efficient source of rubidium
atoms with a minimal impact on UHV conditions led us to the
development of a cold shroud for the rubidium dispenser and MOT
system.  The shroud acts as a pump for rubidium atoms released by
the dispenser that are either (a) not directed towards the MOT, (b)
of the wrong isotope, or (c) moving too quickly to be captured by
the MOT.  The dispenser-shroud system thus acts as a fast, compact,
collimated atomic beam source with a minimized impact on UHV
conditions.

This article discusses the performance of the getter-shroud system
as well as the efficiency of loading atoms into a MOT from the
direct flux of a dispenser. Importantly, our measurements
characterizing the loading rate and equilibrium populations of a MOT
indicate that direct loading of atoms from a getter is strikingly
ineffective.  In contrast, our measurements indicated that a
secondary, lower temperature atomic source was also formed,
contingent on operation of the getter, which was much more effective
at loading a MOT at UHV conditions.  Future getter-loaded, UHV
experiments can be designed to make use of this tempered source in a
more controlled manner.

\section{Experiment}

\begin{figure}
\includegraphics[angle = 0, width =.5\textwidth]{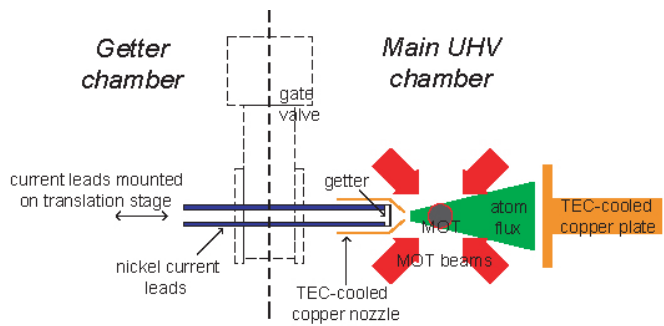}
\caption{\label{exptdiagram} Essential elements of the getter and
cold shroud system. The getter is brought within 1.2 inches of the
MOT center, and with a driving current greater than 2.7 A it
releases a hot rubidium vapor. The cold copper nozzle that surrounds
the getter absorbs nearly all of the emitted rubidium atoms that are
not directed through the open aperture towards the MOT.  The MOT is
loaded from the resultant atomic beam exiting the nozzle.  The vast
majority of emitted atoms are moving too quickly to be captured by
the MOT, but are absorbed by the cold copper plate on the right. The
shroud is comprised of the copper nozzle, the copper plate, and
surrounding mechanical structure.}
\end{figure}

Our experiments are carried out in a stainless steel UHV chamber
pumped to below 10$^{\text{-10}}$ torr.  This chamber was divided
between a main chamber and a secondary ``getter chamber," designed
so as to allow defective or depleted getters to be replaced without
exposure of the main UHV chamber to atmosphere.  The getter is
spot-welded to nickel rods and mounted on a hollow linear
feedthrough \cite{Thermionics:feedthrough} which provides six inches
of travel. The current feedthrough also has a hollow interior (0.4"
ID tube) and a 1.33 inch mini-flange port which accepts an
electrical feedthrough to control the current through the getter. A
single-rod current feedthrough is sufficient, as the return current
path can be grounded to the chamber on the interior bellows of the
linear motion feedthrough.

Figure 1 shows a schematic of the experiment.  The mounted getter on
a retracted translation stage rests just behind a gate valve to the
main chamber.  The vacuum region that surrounds the dispenser and
translation stage is evacuated by a turbomolecular and an ion pump.
Once this region is opened, it takes less than 30 minutes to spot
weld a new getter and reseal the vacuum system.  The ``getter
chamber" is then evacuated and undergoes a modest bakeout
($\approx$2 days), during which the dispenser is degassed as
discussed by the authors of Ref. \cite{fortagh:getter}.  We
typically follow their procedure, although the alternate method
advocated by Ref. \cite{rapol:getter} also provides a usable
rubidium source.

\begin{figure}
\includegraphics[angle = 0, width =
.5\textwidth]{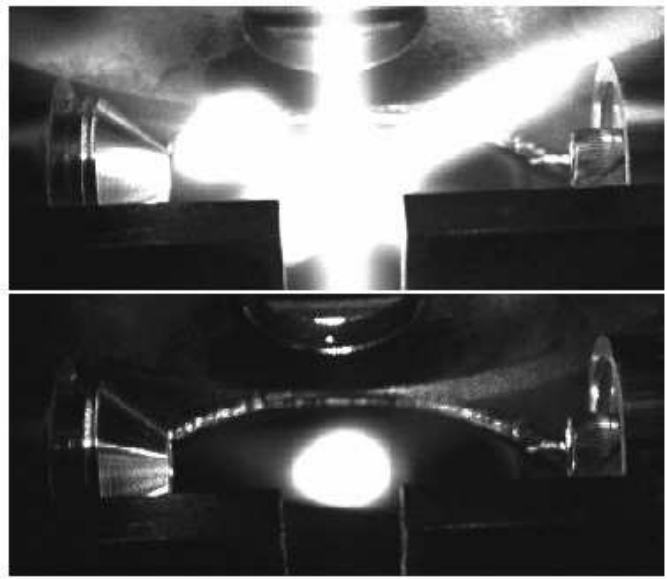}
\caption{\label{roomtempvsbeaming} Extinction of a background vapor
and production of a single-pass atomic beam by low temperature
shroud.  The top image shows the fluorescence of the background
rubidium atoms for a getter pulse of 10 A for 15 seconds at a shroud
temperature of 21.5$^{\text{o}}$C. The bottom image is the
fluorescence for an identical getter pulse and incident MOT beams,
but instead the shroud is held at -30$^{\text{o}}$C.  Note the sharp
edges of the atom beam, as well as the disappearance of background
fluorescence. In both images, the atom beam flows from left to
right, consistent with the orientation of Fig. 1.}
\end{figure}

When the gate valve is opened, pressures in the 10$^{-10}$ torr
range are established in both the getter chamber and the main
chamber.  With the entry path though the gate valve clear, the
dispenser can be translated to within 1.25 inches of the MOT center.

To ensure that rubidium atoms not captured by the MOT are pumped
away, all line of sight from the dispenser to the room temperature
UHV chamber is blocked by a nickel-plated cold copper shroud.  As
the shroud is cooled, the sticking probability for a rubidium atom
(or any alkali metal) impacting the surface approaches unity. In
theory this would protect all sensitive surfaces from the direct
flux of atoms as well as preventing a room temperature background
vapor of rubidium from permeating the chamber during an experiment.
Figure 2 highlights the suppression of the background rubidium vapor
with the cooling of the cold shroud; operating the getter while the
shroud is at room temperature produces visible fluorescence
throughout the chamber due to a thermal rubidium vapor, while
similar operation with a cold (-30 $^o$C) shroud yields fluorescence
only from a beam with line-of-sight access to the getter.  The
pressure spike associated with the dispenser heating, observable at
10$^{\text{-10}}$ torr with an ion gauge, is also reduced by nearly
a factor of two with the cooling of the shroud.

Atoms are collected in a MOT centered within the getter-emitted
beam.  The MOT is formed from 13 mW of laser power to each of six
0.75 inch diameter beams. The quadrupole coils are placed outside
the vacuum system, three inches from the center of the MOT, and are
typically operated with axial field gradients of 20 G/cm. The MOT
population was limited by the relatively small amount of light power
per beam. The trends reported in this paper should scale directly
with improvements in MOT loading.

\begin{figure}
\includegraphics[angle = 0, width =.5\textwidth]{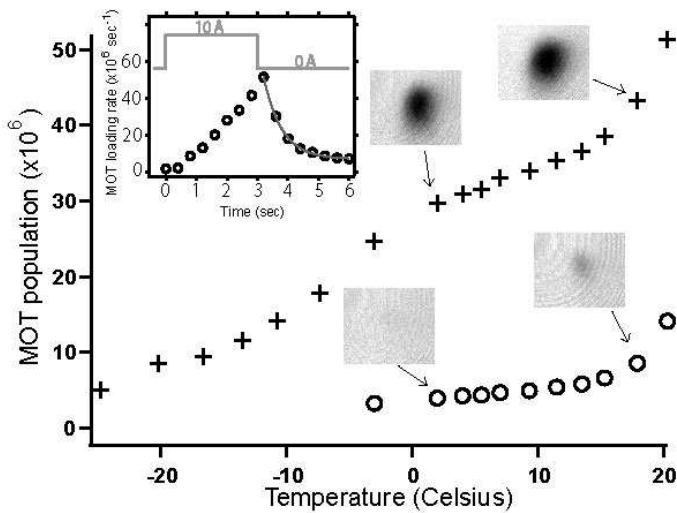}
\caption{\label{MOTvsTemp} MOT population as a function of shroud
temperature. Crosses (+) denote a MOT loaded by a 10 A, 6 sec pulse,
circles (o) a MOT loaded from background atoms for 6 sec with no
getter pulse. Inset shows MOT loading rate as a function of time
overlayed upon the current profile of a 3 sec, 10 A getter pulse
(the shroud temperature was 19$^\circ$C).  The rapid quenching of
the MOT loading rate upon cessation of the getter current has a
decay time of 0.6 sec.}
\end{figure}

\section{Results}

Our pulsed atomic source should be evaluated under two criteria:
first, that operating the source yields an atomic flux which is
efficiently loaded into sizeable MOTs, and, second, that the source
can be quickly switched off to yield UHV conditions for subsequent
experimentation (e.g. magnetic trapping and evaporation) of the
laser-cooled atoms.  We assessed the performance of our
getter-shroud system under both criteria by measuring the loading
rate and equilibrium population in MOTs formed either during or
after the atomic source was pulsed: large MOTs formed during the
getter pulse indicate efficient loading, while small MOTs formed
seconds after the getter pulse indicate the desired suppression of
the atomic flux when the source is turned off. Measurements of MOT
populations under both conditions (getter on/getter off) are shown
in Figure 3 as a function of the temperature of the shroud.

Several conclusions can be drawn from this data.  First, at all
temperatures of the shroud, the MOT populations formed through
operating the getter are much higher than those collected from the
background atomic flux, indicating that the getter-shroud system is
indeed operating as a pulsed atomic beam source, as desired.  Longer
getter pulses make use of this high loading rate to reach
equilibrium MOT numbers up to $3 \times 10^8$.  Second, confirming
the visual findings of Fig. 2, the cold shroud extinguishes the
background rubidium vapor if it is operated at sufficiently low
temperatures; below 0 $^\circ$C, the MOTs formed from the background
vapor were only barely detected by our optical absorption
measurements. Further, the inset of Fig. 3 shows the rapid
termination of the MOT loading rate upon extinguishing the getter
current, satisfying the second stated criterion for the system.

While the getter-shroud system satisfies the stated criteria as an
efficient, rapidly-switched atom source, this data exhibits some
surprising features.  The reason that the getter-loaded MOT depends
so strongly on the shroud temperature, including an almost complete
elimination of trapped atoms for the coldest temperatures, is not
immediately obvious.  The dispenser assembly has no mechanical
contact with the shroud, meaning that the getter itself arrives at
the same temperatures regardless of the temperature of the shroud.
The small MOT populations for low shroud temperatures suggest that
the direct atom flux from the getter is actually a rather poor
source for a MOT. Our original intent was to utilize this shroud at
or below -20 $^o$C, but the reduced MOT population at lower
temperatures forces a choice between rubidium background elimination
and larger MOT populations.  The optimum shroud temperature will
likely vary for different experimental requirements.

In order to diagnose the thermal character of atoms emitted by the
dispenser, a laser 45$^{o}$ to the atomic beam was scanned in
frequency while fluorescence at the center of the MOT region was
detected on a photodiode.  This yields information about the
velocity distribution of the emitted atoms, though the resultant
spectrum is expected to be a convolution of many competing factors
due to the large probe beam size ($\approx$ 0.75 inch diameter),
divergence of emitted atoms, optical pumping rates, and background
fluorescence.

The fluorescence data (Figure 4) show the effect of the cold shroud
on the velocity distribution of the emitted atoms from a strong
getter pulse (10 A, 30 sec). For a shroud at room temperature the
overall rubidium density in the MOT region is approximately twice
that of a -11 $^{o}$C shroud.  At 21 $^{o}$C the background rubidium
vapor and the rubidium flux from the nozzle can be clearly
distinguished, and the Doppler-shifted atoms show peak fluorescence
at 550 MHz from the background vapor. At -11 $^{o}$C, the background
atoms are barely resolvable, and the Doppler-shifted atoms are now
peaked at 630 MHz. Also plotted in Fig. 4 is the fluorescence curve
for a room temperature (21 $^{o}$C) shroud and no getter flux.  In
this case, the peak fluorescence in the MOT region is shifted by 300
MHz from the line center of the Doppler-broadened background
rubidium spectrum.  We believe this is caused by the flux of
desorbing atoms from the shroud nozzle surface which acts as a
directed, room-temperature background flux into the MOT region; we
were only able to discern this fluorescence spectrum after an
atypically large layer of rubidium had been deposited on the inner
surface of the shroud nozzle.

\begin{figure}
\includegraphics[angle = 0, width =
.5\textwidth]{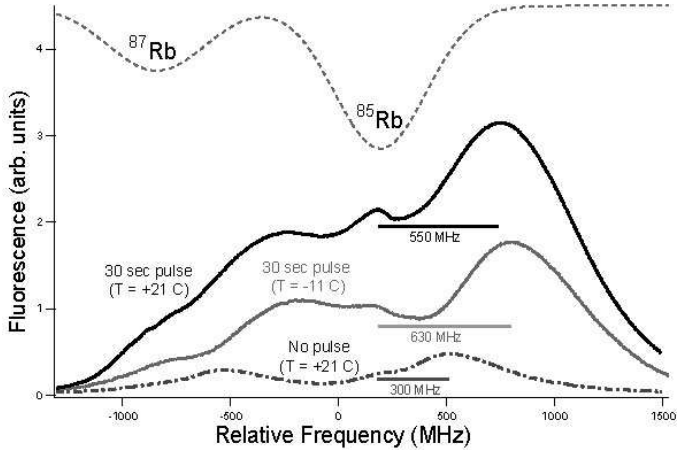} \caption{\label{dopplerBW}
Doppler-sensitive fluorescence spectra of atom flux in the MOT
region. The probe laser beam was incident upon the atom flux at
45$^{o}$ to the center line of the the shroud nozzle.  For
reference, the dotted line shows the room temperature rubidium
spectrum, inferred from an absorption signal generated in a vapor
cell. Plotted are fluorescence curves when the shroud is at room
temperature and at -12 $^{o}$C, given a 30 second, 10 A current
pulse to the getter. Also plotted is the background fluorescence in
the MOT region with no getter pulse, clearly showing the sizeable
room temperature flux emitted from the room temperature shroud
surface after a large layer of atoms have desorbed on the inner
surface of the nozzle.}
\end{figure}

These data explicitly demonstrates that the velocity distribution of
the atomic flux through the nozzle and into the MOT region is
significantly affected by the thermal state of the shroud.  The
strong dependence of the getter-loaded MOT population on shroud
temperature in Fig. 3 is better understood in this context,
especially given the fact that a room temperature beam loads
$\approx$10 times more atoms into a MOT than a 1000 K oven of the
same number flux. The most likely explanation for the modification
of the atom flux is that a large fraction of the atomic flux from
the getter is impingent upon the cold nozzle used to collimate the
remaining flux. If the surface has a substantial probability of
reflecting and \emph{thermalizing} these atoms, then a flux
seemingly colder than the getter-only flux seen at -11 $^{o}$C will
be emitted because the emitted atoms will be a mixture of hot
getter-emitted atoms and cooler reflected atoms. The thermalization
of getter-emitted atoms with surrounding walls has been previously
observed \cite{roach:getter}.  The sum of these two fluxes yields
the reduced Doppler shift seen in the room temperature data in Fig.
4.  We found that after a few weeks of normal operation of the
getter the desorbing flux would then persist for several days.

\section{Discussion}

While the dispenser-shroud system has proven itself to be
potentially useful for ultracold atomic experiments, there are some
drawbacks.  We tried many different avenues for continuous operation
in an attempt to maximize MOT population and minimize the experiment
repetition time.  If the getter is pulsed too frequently ($<$20
seconds separation between 3 sec, 10 A pulses) the pressure in the
chamber rises to a steady state above the minimum base pressure.
While we believe this to be an improvement over bare operation of a
dispenser in UHV, we had hoped to reduce this repetition time
further.  However, even with the coldest shrouds that the TEC system
could effect we were unable to execute an experimental cycle of less
than 20 seconds.

Second, rubidium adsorbed on the shroud will be released into the
UHV if the shroud is allowed to warm up.  If one is to prevent this
substantial gas load from interfering with in-vacuum equipment (such
as ion pumps or high-finesse mirrors), the cooling of the shroud
must be made fail safe.  Furthermore, one would not be able to bake
the shroud in UHV. Thus contamination of the UHV chamber would be
correctable only by selective baking or by thoroughly cleaning the
chamber.  This disadvantage belies the purpose of the getter as an
easily-exchanged UHV atom source.

Drawbacks aside, several improvements to our design could be made
which would make the system a useful tool in many instances. First,
the aperture on the nozzle could be widened to allow a larger flux
into the MOT. The aperture is currently 0.25" in diameter and
appears to ``choke" the MOT at lower temperatures because the beam
is only shining into a fraction of the MOT cross-section. A larger
aperture would increase the flux as the square of the aperture
diameter, allowing the maximum getter-loaded population number to be
reached in less time with a reduced impact on the pressure.

Another improvement would be the addition of a ``shadow" for the
getter. This would likely take the form of a metal piece which
would obscure the getter slit from direct line of sight to the
center of the MOT.  This should drastically reduce the losses due
to MOT atoms colliding with fast Rb atoms, allowing for a larger
final MOT population.

Finally, given our understanding of the fluorescence spectrum in
Fig. 4, one could construct a system which would utilize the
secondary room temperature beam generated by the rubidium-coated
surfaces of the shroud.  A miniature ``oven," operated in the main
UHV chamber, could surround a dispenser that purposefully directs
its atomic flux towards the inner walls of the oven.  Atoms emitted
from the getter would be tempered by the inner surface of the oven,
and these thermalized atoms would then be allowed to escape through
a collimated aperture to efficiently load a closely situated MOT.
Cold baffles would then be placed behind the MOT to pump away the
atoms which are not captured by the MOT.  During experimental
operation the oven could be held at or slightly above room
temperature to increase atom yield and prevent a surface layer of
atoms from forming.  When the experimental system is not in
operation the miniature oven could be heated further to release any
remaining adsorbed atoms onto the cold baffles.

\begin{acknowledgments}

We thank Stefan Schmid, Mike Grobis, and Dave Murai for their
technical assistance and advice. The authors' effort was sponsored
by the Defense Advanced Research Projects Agency (DARPA) and Air
Force Laboratory, Air Force Materiel Command, USAF, under Contract
No. F30602-01-2-0524, the National Science Foundation under Grant
No. 0130414, the Sloan Foundation, the David and Lucile Packard
Foundation, and the University of California.  KLM acknowledges
support from the National Science Foundation. SG acknowledges
support from the Miller Institute for Basic Research in Science.

\end{acknowledgments}

\bibliographystyle{prsty}

\begin{thebibliography}{10}
\bibitem{Thermionics:feedthrough}
Manufactured by Thermionics Vacuum Products, part number
FLMH-275-50-6.  Inclusion of this item is not an endorsement; we
expect that devices from other manufacturers would behave in a
similar fashion.
\end{thebibliography}

\end{document}